\documentclass[a4paper,11pt]{article}
\pdfoutput=1
\usepackage{jcappub}
\usepackage{subfig}
\usepackage{epsfig}
\usepackage{graphicx}
\usepackage{color}
\usepackage{amssymb}
\usepackage{amsmath}
\usepackage{mathtools}
\usepackage{url}
\usepackage{natbib}
\usepackage[normalem]{ulem}
\usepackage{afterpage}
\usepackage{float}

\newcommand{\bfe}{{\mathbf e}}

\newcommand{\by}{{\mathbf y}}
\newcommand{\bx}{{\mathbf x}}

\newcommand{\bn}{{\mathbf n}}

\newcommand{\bel}{{\boldsymbol \ell}}

\newcommand{\HH}{{\cal H}}

\newcommand{\cd}{\cdot}

\newcommand{\br}{{\bf r}}

\newcommand{\de}{\delta}
\newcommand{\De}{\Delta}

\newcommand{\ga}{\gamma}

\newcommand{\ka}{\kappa}

\newcommand{\La}{\Lambda}

\newcommand{\Om}{\Omega}

\newcommand{\vth}{\vartheta}

\newcommand{\ra}{\rightarrow}

\newcommand{\be}{\begin{equation}}
\newcommand{\ee}{\end{equation}}
\newcommand{\gsim}{\stackrel{>}{\sim}}

\newcommand{\bea}{\begin{eqnarray}}
\newcommand{\eea}{\end{eqnarray}}
\newcommand{\bean}{\begin{eqnarray*}}
\newcommand{\eean}{\end{eqnarray*}}
\newcommand{\dd}{\partial}

\newcommand*\bra[1]{\left(#1 \right)}

\newcommand\spart{\;\raise1.0pt\hbox{$/$}\hskip-6pt\partial}
\definecolor{dred}{rgb}{0.8,0,0.1}

\title{General Relativistic corrections in density-shear correlations}

\author[]{Basundhara Ghosh,}
\author[]{Ruth Durrer,}
\author[]{Elena Sellentin}
\affiliation[]{D\'epartement de Physique Th\'eorique and Center for Astroparticle Physics,\\
Universit\'e de Gen\`eve, 24 quai Ernest  Ansermet, 1211 Gen\`eve 4, Switzerland}

\emailAdd{basundhara.ghosh@unige.ch}
\emailAdd{ruth.durrer@unige.ch}
\emailAdd{elena.sellentin@unige.ch}

\abstract{
We investigate the corrections which relativistic light-cone computations induce on the correlation of the tangential shear with galaxy number counts, also known as galaxy-galaxy lensing. The standard-approach to galaxy-galaxy lensing treats the number density of sources in a foreground bin as observable, whereas it is in reality unobservable due to the presence of relativistic corrections. We find that already in the redshift range covered by the DES first year data, these currently neglected relativistic terms lead to a systematic correction of up to 50\% in the density-shear correlation function for the highest redshift bins. This correction is dominated by the the fact that a redshift bin of number counts does not only lens sources in a background bin, but is itself again lensed by all masses between the observer and the counted source population. Relativistic corrections are currently ignored in the standard galaxy-galaxy analyses, and the additional lensing of a counted source populations is only included in the error budget (via the covariance matrix). At increasingly higher redshifts and larger scales, these relativistic and lensing corrections become however increasingly more important, and we here argue that it is then more efficient, and also cleaner, to account for these corrections in the density-shear correlations.}


\arxivnumber{}

\begin{document}
\maketitle
\flushbottom

\section{Introduction}
Currently  one of the most impressive success stories in cosmology is the highly accurate observation and our detailed understanding of the cosmic microwave background (CMB), its anisotropies and its polarisation, see e.g.~\cite{Ade:2013ktc,Planck:2015xua,Ade:2015tva,Durrer:2015lza,Durrer:2008aa}. As a cosmological community, we would now like to repeat this success story at lower redshifts by using present and future galaxy surveys. Contrary to the CMB which primarily comes from the two-dimensional surface of last scattering, galaxy surveys are three-dimensional and therefore contain  more, potentially richer information. Especially by using tomography, i.e., by splitting a source population into different redshift bins, we can study how cosmic structure formation proceeds and thereby directly test the gravitational instability picture.

To repeat the CMB success story, it is important that we make optimal use of the low-redshift data. Of course, the fact that clustering becomes non-linear on smaller scales and late time as well as the influence of non-gravitational hydrodynamical effects and more, render the interpretation of the data more difficult. The current standard-approach to model these non-linearities is to translate a linear power spectrum to a non-linear power spectrum by using `Halofit' \citep{Takahashi:2012em}, and in this paper we shall adhere to this approach, noting however that it only refers to the non-linear growth of scalar perturbations in Newtonian gravity. General Relativistic N-body codes do exist and the agreement of the matter power spectrum from relativistic simulations with Halofit is excellent~\cite{Adamek:2017uiq}.

The problem of galaxy formation depending on its environment is on an observational level modeled by `biasing': the observed galaxy density distribution $\delta_g$ is assumed to be related to the underlying matter density distribution, $\de_g=b(z)\de$ via some biasing function $b$. When cosmological parameters are inferred, this bias function is treated as nuisance parameters, and marginalized over. Another nuisance parameter in shear correlations is the intrinsic alignment contribution to the shear signal. Also this is  marginalized over. The free parameters of these bias functions are likely to mimic, to some extent, the contribution from relativistic effects which we study in this paper. However, since these effects are signals of theoretical interest which can be calculated, marginalizing over them is suboptimal.

In this paper we  investigate the impact of general relativistic corrections on the correlation between number counts in a foreground bin, and lensing in a background bin. Our starting point is that it has lately been shown~\cite{Yoo:2009au,Yoo:2010ni,Bonvin:2011bg,Challinor:2011bk} that counting galaxies in a fixed solid angle and redshift bin does not directly measure the galaxy over-density. The resulting count is not only affected by redshift space distortions~\cite{Kaiser1987}, but also enhanced or decreased by lensing and magnification bias~\cite{Matsubara:1999du} and by large-scale relativistic effects. The relativistic effects other than lensing are mainly relevant on very large scales where they can mimic a primordial non-Gaussianity~\cite{Camera:2014bwa,Raccanelli:2015vla}. Redshift space distortions are well known and are routinely used to constrain the cosmic growth factor~\cite{Vipers2017,Vipers2017b}.

In recent years it has been shown that also the lensing and magnification bias term is considerable and will be measured in future galaxy clustering analyses~\cite{Montanari:2015rga}. Furthermore, neglecting it can lead to misinterpretation of results from galaxy surveys, see, e.g.~\cite{Cardona:2016qxn,Dizgah:2016bgm,Villa:2017yfg}.
Also in this paper we investigate the effects of the lensing term. More precisely, we focus on galaxy-galaxy lensing, where the tangential shear of background sources is correlated with number counts in the foreground. We study how relativistic effects, and especially the lensing term from projection, affect the correlation of galaxy density fluctuations with the tangential shear. We find, that already in present surveys the effect can contribute up to 50\%
at high redshift. That this term has to be included in the galaxy-galaxy lensing cross correlation is of course not new. It has already been discussed ten years ago in Ref.~\cite{Ziour:2008iz} and subsequently in several other papers. Nevertheless it is not included in present analyses. Here we present a concrete case study to estimate the maximum size of the effect in recent surveys.

This paper is structured as follows: In the next section we derive the theoretical expressions for the cross-correlation function between relativistically correct number counts and the tangential shear, and its corresponding angular power spectrum in redshift space. In Section~\ref{s:num} we present some numerical examples and discuss them. We show especially how the recently published DES (Dark Energy Survey) first year results~\cite{Abbott:2017wau} could gain in precision by including the lensing term directly into the signal of the number counts. In Section~\ref{s:con} we conclude. We present a comprehensive derivation of the density-shear correlation function in the full and flat sky in Appendix~\ref{a:der}.

\section{Correlating number counts with shear measurements}
Here we present the first order expression for the correlation between galaxy number counts and the tangential shear. For completeness, a detailed derivation is presented in Appendix~\ref{a:der}, where we also make the connection to results known from the literature.

In first order perturbation theory, the number of galaxies within a redshift bin $dz$ and a solid angle $d\Om$ at observed redshift $z$,  in observed direction $\bn =\bn(\vth,\varphi)$ is given by
\bea 
N(z,\bn)=\bar N(z)[1+\De(z,\bn, m_{\rm lim})]
\eea
where $ \bar{N}(z)$ is the average spatial number density at redshift $z$, and where the observable over-density is~\cite{Bonvin:2011bg,Challinor:2011bk,DiDio:2013sea,Camera:2014bwa}
\bea
\De(\bn,z,m_{\rm lim}) &=& b(z)\de +\frac{1}{\HH}
\left[\dot\Phi+\dd^2_rV\right]  +(2-5s)\left[\int_0^{r}\hspace{-0.3mm}\frac{d\tilde r}{r} (\Phi+\Psi) - ~ \ka\right] +(f_{\rm evo}-3)\HH V  +   \nonumber \\  &&
(5s-2)\Phi + \Psi+ \left(\frac{{\dot\HH}}{\HH^2}+\frac{2-5s}{r\HH} +5s-f_{\rm evo}\right)\left(\Psi+\dd_rV+
 \int_0^{r}\hspace{-0.3mm}d\tilde r(\dot\Phi+\dot\Psi)\right)  \,.
   \nonumber \\  &&
  \label{e:DezNF}
\eea
Here an overdot denotes a derivative with respect to conformal time, $\HH$ is the conformal Hubble parameter and $r=r(z)$ is the comoving distance to redshift $z$. The peculiar velocity is given by $V$, the velocity potential in longitudinal gauge, such that velocity components are given by $v_i=-\dd_iV$. The term  $\de$ is the matter density fluctuation in comoving gauge: on small scales it reduces to the Newtonian density contrast, but it is by itself not observable, even if the galaxy bias function $b(z)$ were known. The quantities $\Phi$ and $\Psi$ are the Bardeen potentials. More details on Eq.~(\ref{e:DezNF}) are given in~\cite{DiDio:2013sea} and~\cite{DiDio:2013bqa}. 

Furthermore,  denoting the angular Laplacian as $\Delta_\Om$, the convergence $\ka$ is given by the angular Laplacian of the lensing potential, $\phi$,
\bea\label{e:kappa}
\ka &=& -\frac{1}{2}\Delta_\Om\phi \,, \\
\phi(\bn,z) &=& -\int_0^{r(z)}\!\!d\tilde r\frac{r(z)-\tilde r }{r(z)\tilde r}(\Phi+\Psi)(\tilde r\bn,\tau_0-\tilde r)\,. \label{e:lenpot}
\eea
We denote the limiting luminosity by $L_{\rm lim}$. The evolution bias, $f_{\rm evo}$, captures the fact that new galaxies form and galaxies merge as the Universe expands and hence their number density evolves not simply as $(1+z)^{3}$;  $f_{\rm evo}$ depends on redshift and on $L_{\rm lim}$ and is defined as
\be
\label{eq:fevo}
f_{\rm evo}(z,L_{\rm lim}) \equiv \frac{\partial\ln\bra{a^3 \bar N(z,L>L_{\rm lim})}}{ \partial \ln a} \, .
\ee
Here $\bar N(z,L> L_{\rm lim})$ is the background number density of galaxies with luminosity above $L_{\rm lim}$ and $a=1/(z+1)$ is the cosmic scale factor.
Finally, we introduce magnification bias: due to magnification, less luminous galaxies still make it into our survey if they are in a region of high magnification and vice versa. Denoting the  limiting magnitude of the survey
$m_{\rm lim}$, the magnification bias is given by
\be
s(z,m_{\rm lim}) \equiv \left.\frac{\partial\log_{10}{\bar N}(z,L> L_{\rm lim})}{\partial m}\right|_{m_{\rm lim}} \,.
\label{e:s_mlim}
\ee
The redshift dependence of this quantity depends on the specific survey. Only if we see all galaxies of a considered type, i.e. if the survey is complete, we have $s=0$. Note that $s$ nearly always enters in the combination $5s-2$. This is the factor which multiplies the fluctuations of the angular diameter (area) distance~\cite{Bonvin:2005ps}, $\de D_A(z)$  which contributes twofold: it leads to an increase in the transversal volume (area) and hence to a decrease in the density; this is the term $-2$. But it also increases the observed brightness in sources of a given luminosity and can bring them into an incomplete survey, enhancing the density; this is the term $+5s$ which is also called magnification bias. In the combination $(5s-2)\ka$ we shall call them here the ``lensing contribution" to the actually observable $\De(\bn,z)$.

In Appendix~\ref{a:der} we derive the following expression for the correlation function between the observable galaxy number count $\De$ and the tangential shear $\ga_t$
\be
\langle \De(\bn,z)\ga_t(\bn',z') \rangle = 
\frac{-1}{4\pi}\sum_\ell\frac{2\ell+1} {\ell(\ell+1)}P_{\ell\,2}(\bn\cd\bn')C_\ell^{\De,\ka}(z,z')
 \,.
\ee
Here, $P_\ell m(\mu)$ is the Legendre polynomial of degree $\ell$ and order $m$, and $C_\ell^{\De,\ka}(z,z')$ is the angular power spectrum of the correlation between $\De$ and $\ka$. We split it into its correlation with the not-directly observable density contrast (`$\delta$'), augmented by redshift space distortions (`rsd'), by the convergence $\ka$ and by large scale relativistic effects (`ls')
\be\label{e:comp}
C_\ell^{\De,\ka}(z,z') = b(z)C_\ell^{\de,\ka}(z,z') + C_\ell^{{\rm rsd},\,\ka}(z,z')-(2-5s(z))C_\ell^{\ka,\ka}(z,z') + C_\ell^{{\rm ls},\,\ka}(z,z') \,.
\ee
Here, $C_\ell^{{\rm rsd},\ka}$ denotes the correlation of $\ka$ with the redshift space distortion (rsd) term caused by $\dd_r^2V$, while $C_\ell^{{\rm ls},\ka}$ denotes its correlation with all the remaining relativistic terms which are relevant mainly on very large scales. They also include the so-called Doppler term $\propto \dd_rV$ which is strictly speaking not relativistic but also only relevant on large scales. The lensing term $-(2-5s(z))C_\ell^{\ka,\ka}$ and  the large scale corrections, $ C_\ell^{{\rm ls},\,\ka}$, apart from the Doppler term are due to General Relativity and we call them `General Relativistic corrections'. We shall see that on intermediate and small scales, $\ell>20$, the lensing term by far dominates these corrections to the number density.

In standard analyses, which includes the DES analysis, only the term $C_\ell^{\de,\ka}(z,z')$ is considered. The main point of this paper is to show that this introduces systematic deviations in the signal and that in particular the lensing of the number densities in the foreground bin cannot be neglected.
In the next section we compute all the terms for several examples and we show that only the $C_\ell^{\ka,\ka}(z,z')$ term is a relevant correction for present and near future galaxy clustering surveys. Redshift space distortions are always much smaller and the relativistic terms contribute only on very large scales, $\ell<10$, where cosmic variance is significant and which are not accessed in the DES survey.

\section{Numerical Examples}\label{s:num}
In this section we present changes in the signal for  numerical evaluations of the above relativistic contributions for the standard $\La$CDM cosmology. We use the public code {\sc class}~\cite{Lesgourgues:2011re,Blas:2011rf} which has been expanded to include the relativistic contributions to galaxy number counts~\cite{DiDio:2013bqa}. We assume purely scalar perturbations with cosmological parameters of the Planck-2015 results~\cite{Planck:2015xua}. More precisely, we set the Hubble parameter $H_0=67.556$~km s$^{-1}$ Mpc$^{-1}$, the baryon density parameter $\Omega_bh^2=0.022032$, the cold dark matter density parameter $\Omega_{\rm cdm}h^2=0.12038$, the curvature $K=0$, the number of neutrino species $N_\nu=3.046$ and  the neutrino masses are neglected.  We set the galaxy bias to unity, $b(z)=1$, and assume a complete survey, $s=0$, for our analysis.

We first present results for a mock survey which mimics the redshift binning of the first year DES observations. 
A similar analysis with somewhat smaller sky coverage has also been published by KiDS (Kilo Degree survey)~\cite{vanUitert:2017ieu}. In contrast to DES, we assume however full-sky coverage since our interest also includes the very low multipoles where relativistic effects leave noticeable traces.

In this paper we do not want to exactly determine the contribution to the DES correlation functions and error budget from our terms, we just want to give the correct order of magnitude. Clearly, since this is a systematic effect which is easy to model, it would be more useful to add it to the data than to just include it in the error budget.

Furthermore, if $2-5s(z)>0$, which is true in observed volumes where a survey reaches near completeness, then the $C_\ell^{\ka,\ka}$ contribution is negative as is $C_\ell^{\de,\ka}$ so that $|C_\ell^{\de,\ka}-2 C_\ell^{\ka,\ka}| >|C_\ell^{\de,\ka}| $.

\subsection{A generic survey with DES-like redshift binning}
The DES collaboration has presented first year data on the cross-correlation of galaxy clustering and lensing ~\cite{Abbott:2017wau}. This analysis uses five galaxy redshift bins in the foreground (abbreviated by `f') with width $\De z=0.15$. These foreground populations are correlated with four tangential shear redshift bins in the background, (abbreviated by `b') with widths $\De z=0.23,~0.2,~0.27$ and $0.23$ respectively. The mean redshifts of these bins are given by
\bea
z_{f1}= 0.225\quad z_{f2}= 0.375\quad z_{f3}=0.525 \quad z_{f4}=0.675 \quad  z_{f5}=0.825\\
z_{b1}= 0.315\quad z_{b2}=0.53 \quad z_{b3}=0.765 \quad z_{b4}=1.1 \quad 
\eea
In order to establish whether relativistic corrections have a noticeable impact on current surveys, we model a generic DES-like survey by using the above redshifts for centers of Gaussian redshift bins with the corresponding widths. If relativistic effects have a noticeable impact, then the correlations $C_\ell^{\de,\ka}(z_{fi},z_{bj})$, where only  $\delta$-terms are included, deviate from the correlations $C_\ell^{\De,\ka}(z_{fi},z_{bj})$, where also the relativistic terms are included.

\begin{figure}[H]
\centering
  {\includegraphics[width=0.8\textwidth]{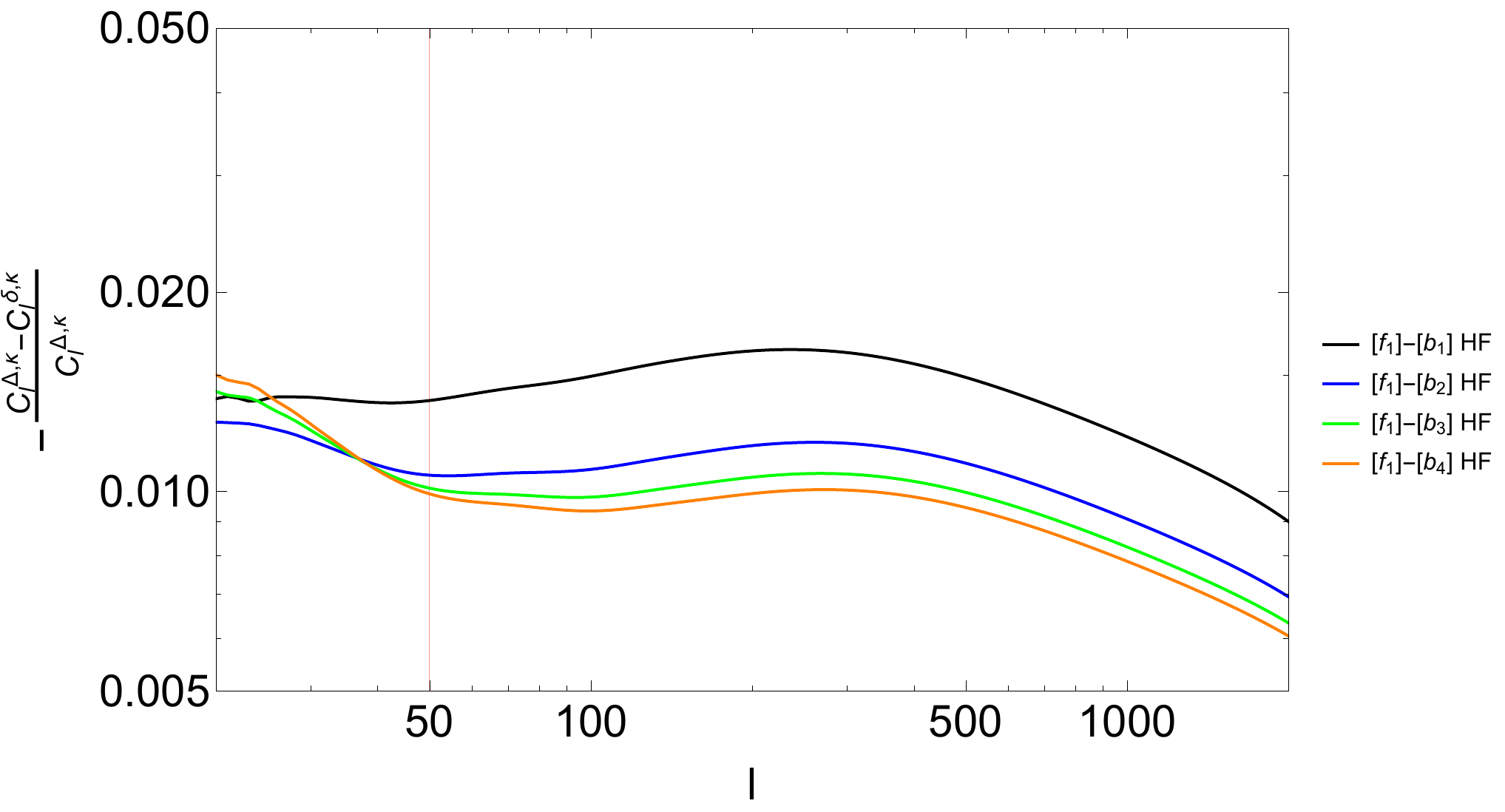}} \\
\hspace{0.45cm}  {\includegraphics[width=0.807\textwidth]{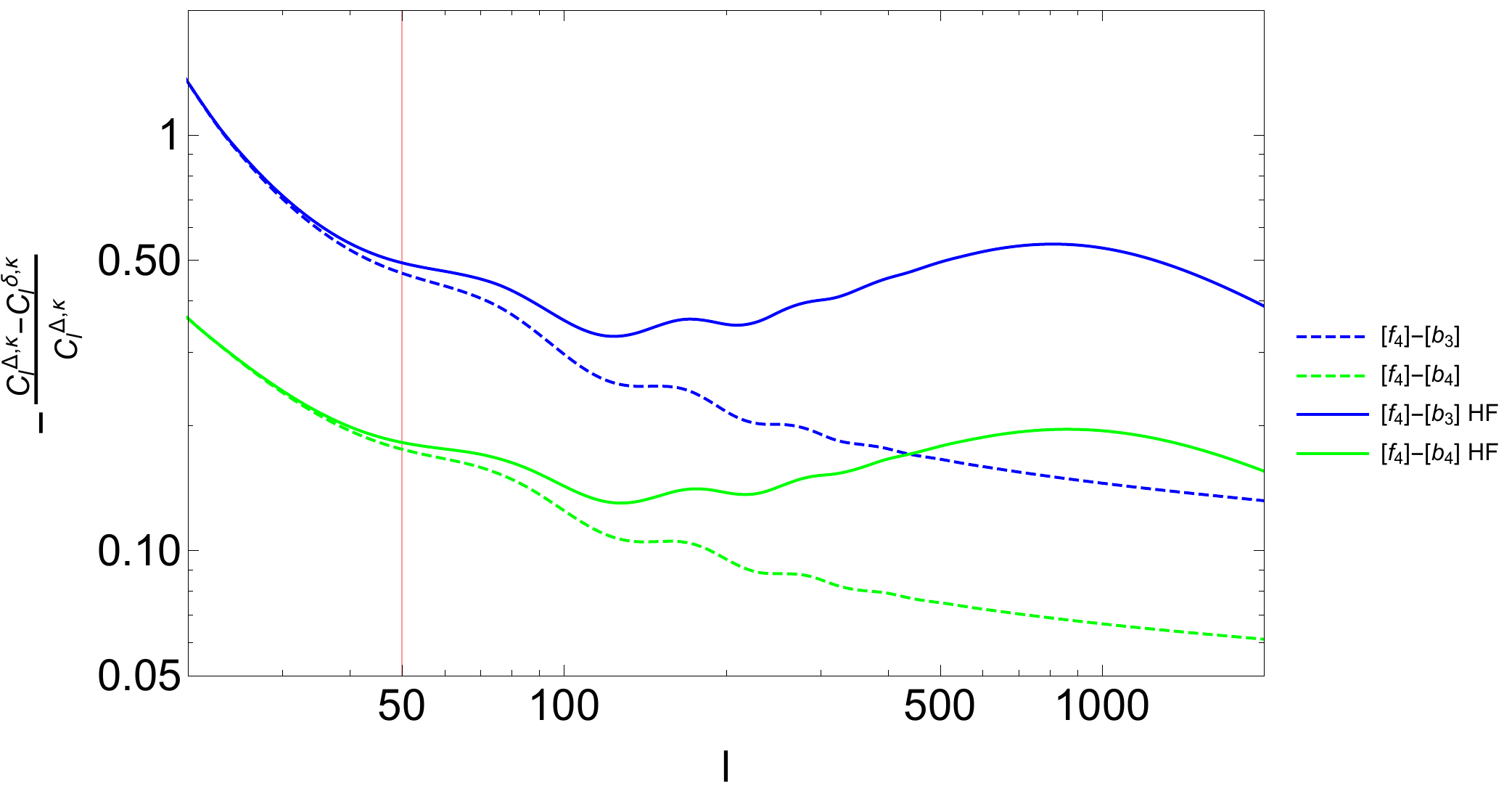}}
{\includegraphics[width=0.8\textwidth]{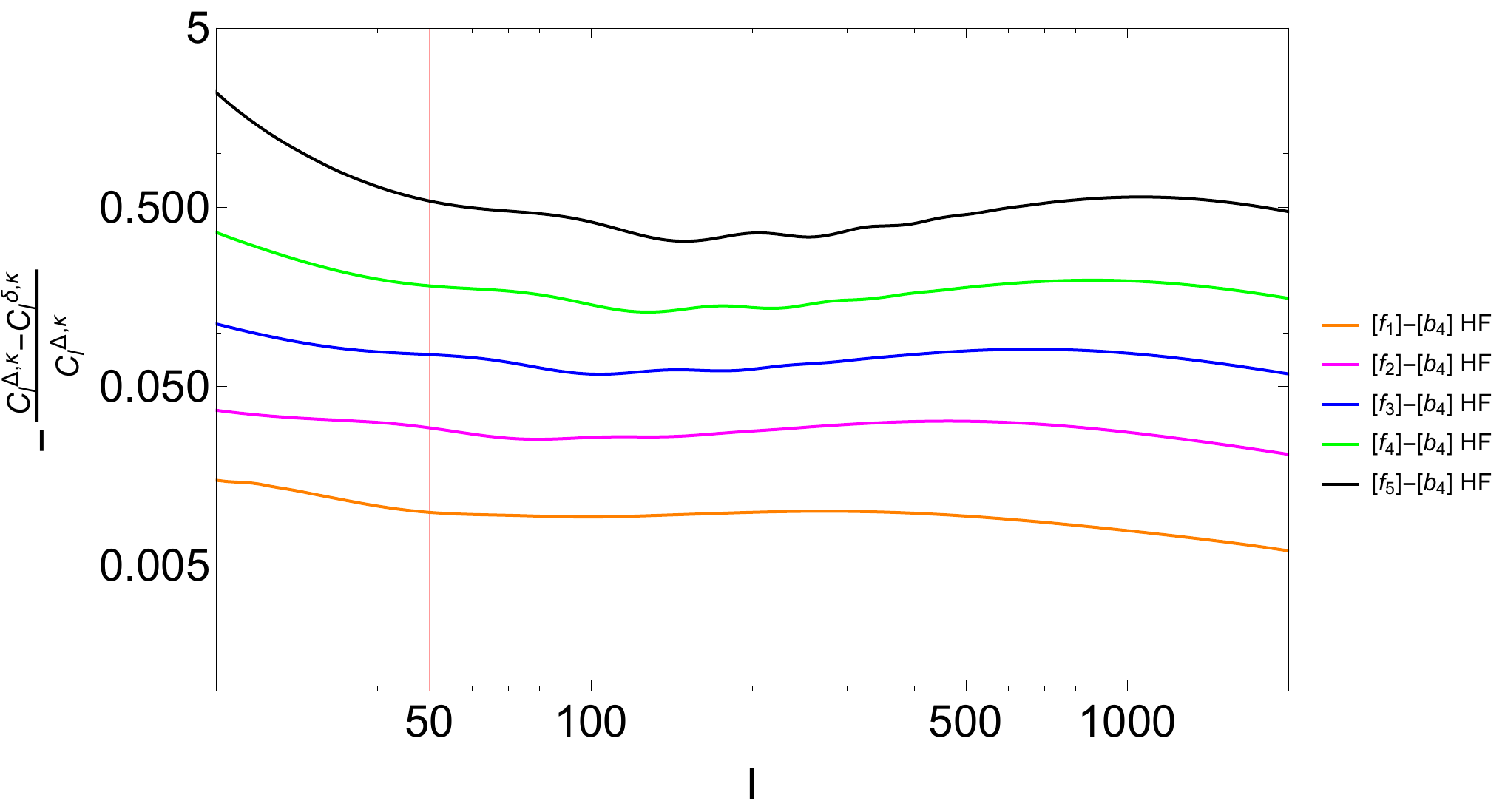}}
\caption{The top and middle panels show the relative contribution of all terms apart from $\delta$ for the cross-correlations between the lowest foreground bin and second highest foreground bin denoted here by $[f_1]$ and $[f_4]$ respectively, with the higher background bins. This is expressed in terms of the relative difference in angular power spectra of the mentioned contributions. The bottom one is for all foreground bins correlated with the highest background $[b_4]$. HF stands for Halofit. DES results are only sensitive to the values of $\ell$ above the vertical red line at  $\ell=50$. The signals are all negative, and hence the vertical axis is written with a negative sign. }
 \label{f:delta}
\end{figure}

\begin{figure}[H]
  \centering
  {\includegraphics[width=0.83\textwidth]{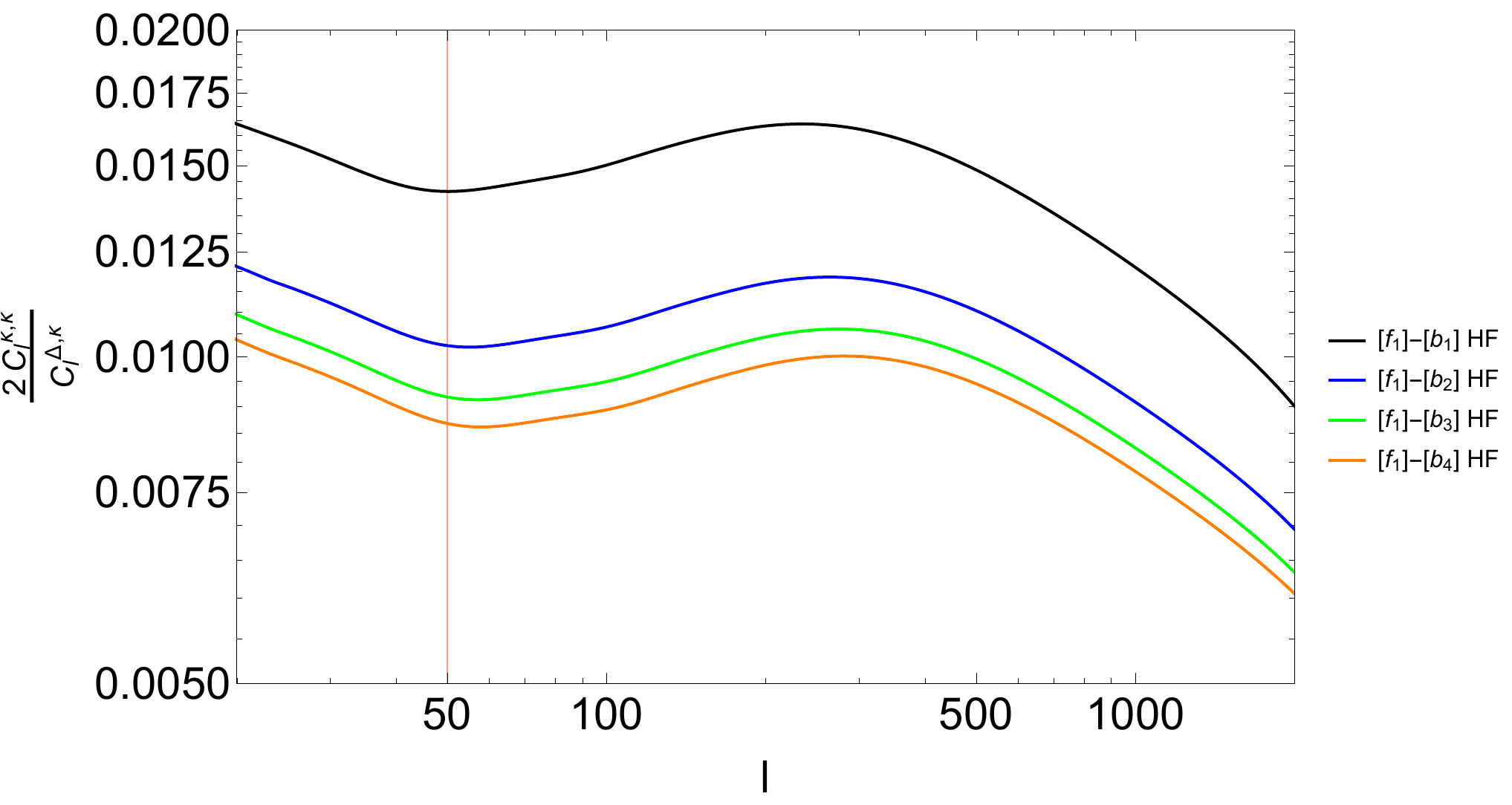}} \\
\hspace{0.45cm}
  {\includegraphics[width=0.8307\textwidth]{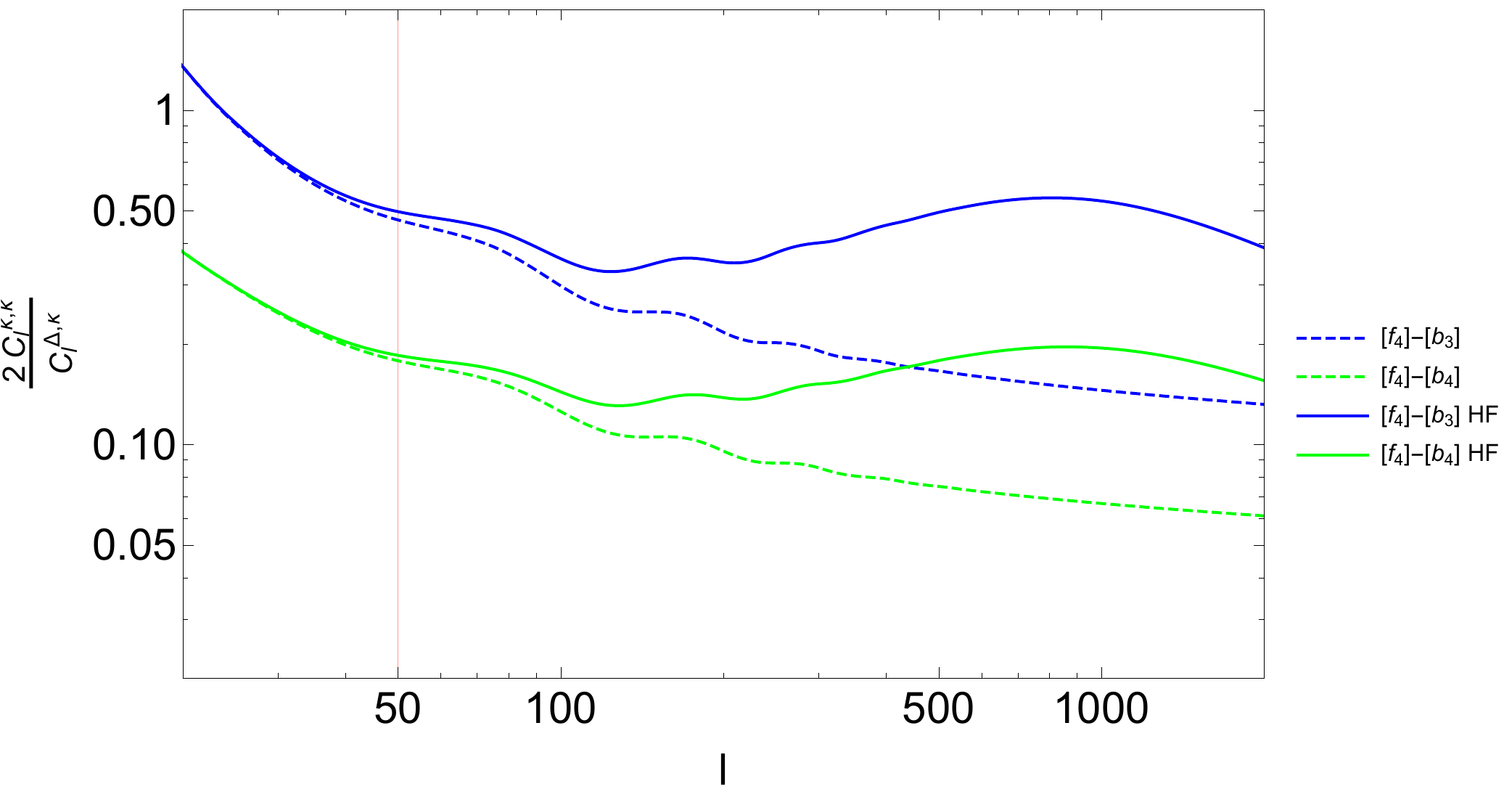}}
{\includegraphics[width=0.83\textwidth]{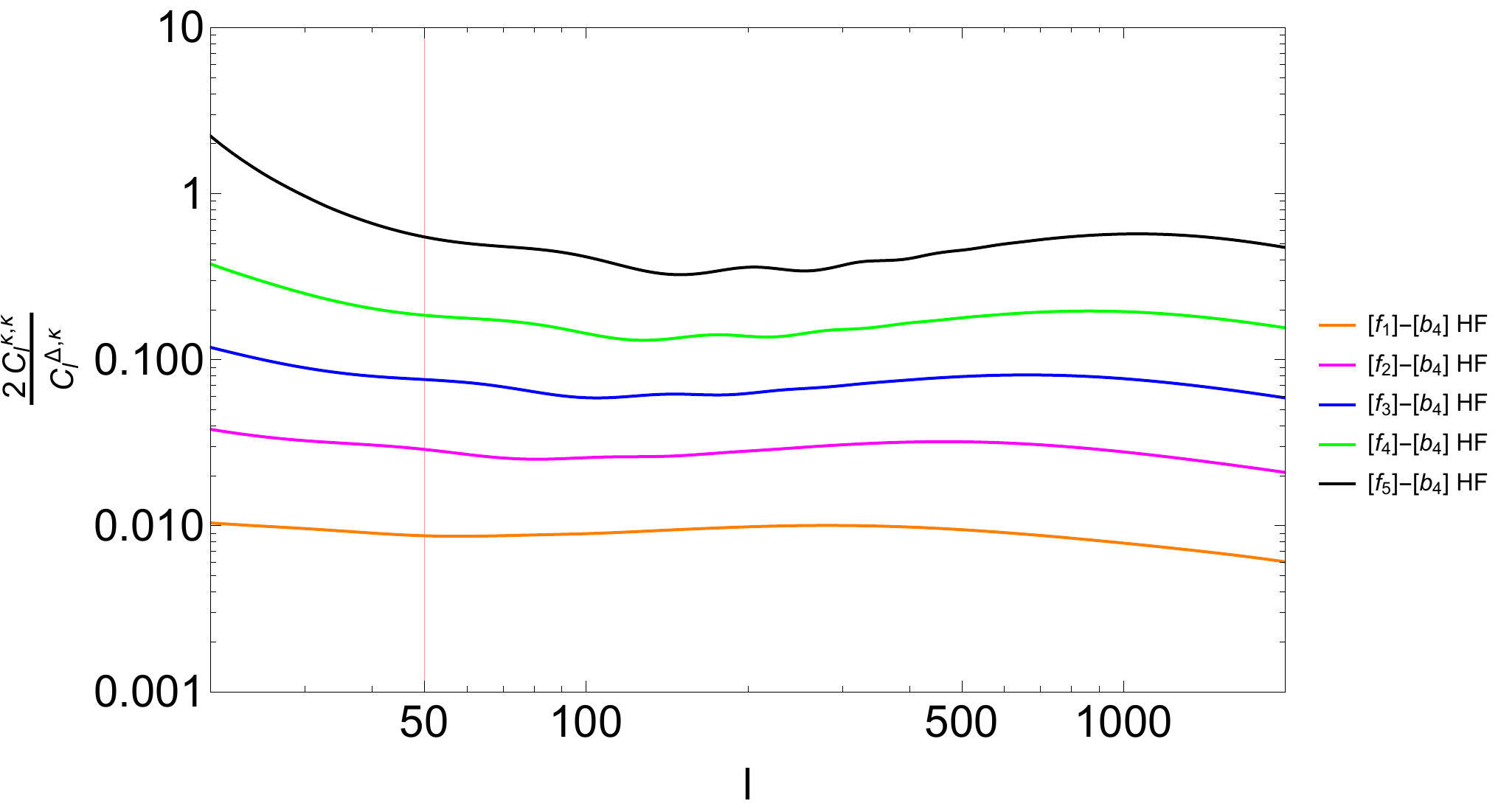}}
\caption{The same as in Fig.~\ref{f:delta} but here  $C_\ell^{\De,\ka}-C_\ell^{\de,\ka}$ is replaced by $-2C_\ell^{\ka,\ka}$.}
 \label{f:kappa}
\end{figure}

We have thus calculated the contribution of $C_\ell^{\de,\ka}(z_{fi},z_{bj})$ in $C_\ell^{\De,\ka}(z_{fi},z_{bj})$ for all $z_{fi}<z_{bj}$.
 In Fig.~\ref{f:delta} we show the full relative difference  $(C_\ell^{\De,\ka}-C_\ell^{\de,\ka})/C_\ell^{\De,\ka}$, whereas in Fig.~\ref{f:kappa} we show the relative contribution $2C_\ell^{\ka,\ka}/C_\ell^{\De,\ka}$ from the convergence $\kappa$ alone. 

As it can be seen in the top panel of Fig.~\ref{f:delta}, for the lowest foreground bin, the subdominant terms apart from $\delta$ (as seen in Eq.~\ref{e:comp}) contribute up to 1.5\% to the total on large scales ( $\ell>20$). The scales accessible to the DES 1st year survey correspond to $\ell>50$, where the contribution of the subdominant terms is small, around 1-1.5\%. For $\ell\gsim 40$, lower redshift lensing bins which are closer to the foreground source always dominate while for $\ell<40$ this is no longer so. For very wide angles the relativistic effects which increase for larger redshifts contribute significantly.

In the bottom panel we show the relative contribution from the `subdominant' terms for the highest background bin and all forground bins. Of course this is a monotonically increasing function of the foreground redshift. It rises up to 50\% for $\ell>50$ which contribute to the  DES results. Clearly, these contributions cannot be neglected. 
Note that the results obtained in the top and bottom panel of  Fig.~\ref{f:delta} have been obtained using `Halofit'~\cite{Takahashi:2012em} for the density power spectrum. This is relevant above $\ell\sim 50$ as is shown in the middle panel where the results with (dashed) and without (solid) Halofit are compared for two redshift bin combinations.

In Fig.~\ref{f:kappa} we now show simply $-2C_\ell^{\ka,\ka}/C_\ell^{\De,\ka}$. We have verified that for $\ell\gsim 50$ this is virtually identical to $(C_\ell^{\De,\ka}-C_\ell^{\de,\ka})/C_\ell^{\De,\ka}$, hence all the difference is  actually due to lensing. It ranges from a mere 1\% for the lowest foreground bin to more than 50\% for the highest one, which assures us that the contribution will increase further with the higher redshift bins of future surveys (see also Fig.~\ref{f:highz}).
At low $\ell$ the lensing contribution alone does not explain all the signal and redshift space distortions and relativistic effects can become significant.

\begin{figure}[]
\centering
{\includegraphics[width=\textwidth]{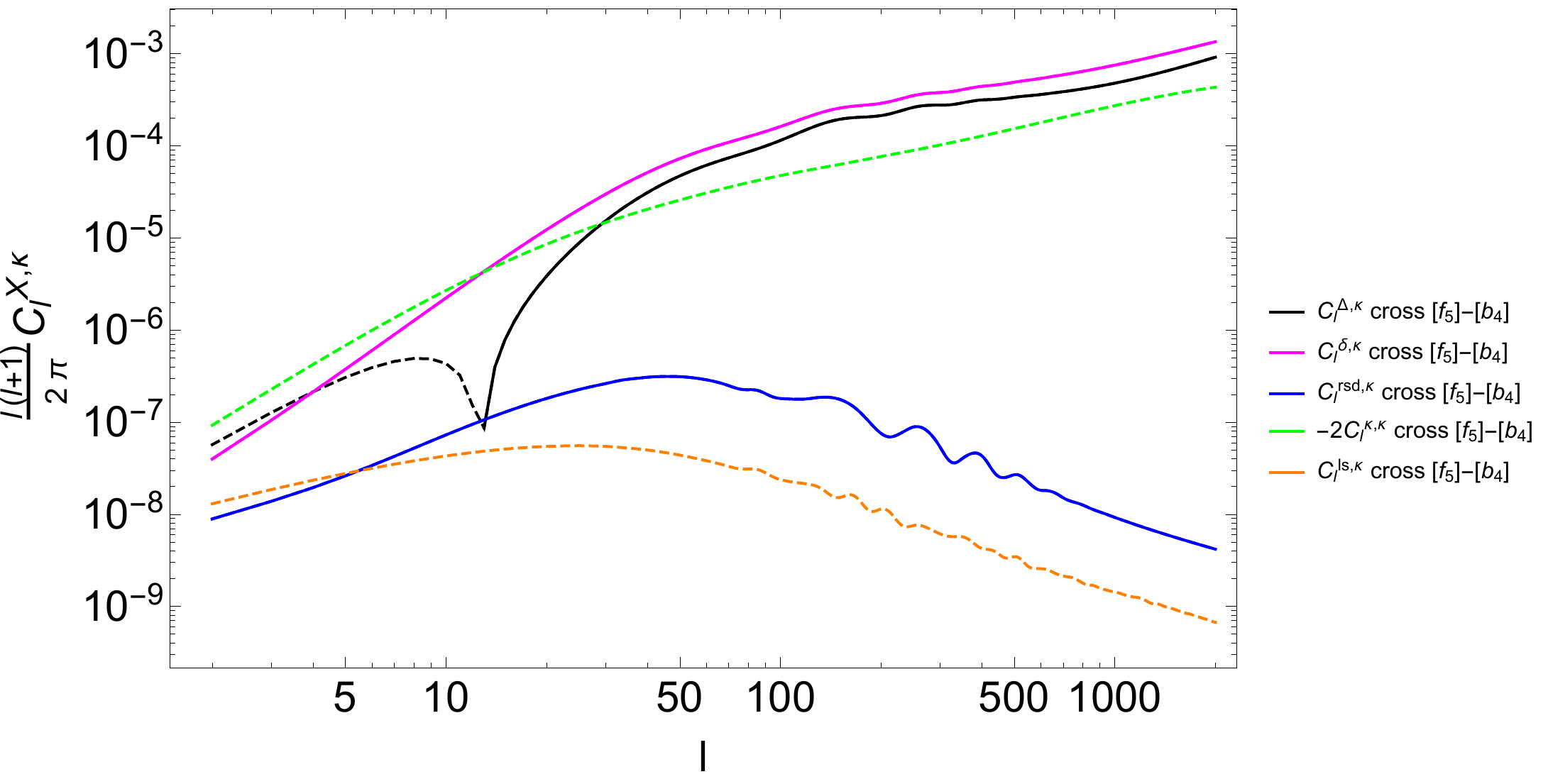}}
\caption{This figure  compares  the different contributions to the number count shear cross-correlation, always using Halofit. The black line represents the total galaxy number counts--tangential shear correlation, the magenta, green, blue and orange ones are for the density, $\ka$, redshift space distortion and large scale relativistic effects respectively. The dashed lines indicate negative signals.}
\label{f:compare}
\end{figure}

 In Fig.~\ref{f:compare}, we show and compare the different contributions, with the total galaxy number count signal shown in black. This is done only for the highest redshift bins, $z_{f5}$ and $z_{b4}$, since the differences are most prominent in this case. While we see that the redshift space distortions and other relativistic effects have comparatively small contributions, the one due to lensing is definitely important, it is in fact larger than the density term at very large scales, $\ell<10$.
For lower redshifts, the lensing term is smaller and we have found that at low $\ell$ redshift space distortions and relativistic effects cannot be neglected. 

For a comparison with the DES first year results only the $C_\ell$'s with $\ell>50$ are relevant, since the sky coverage of the DES first year is about $1/30$th of the full sky.

\subsection{Future surveys and higher redshifts}
\begin{figure}[!htb]
 \centering
{\includegraphics[width=0.8\textwidth]{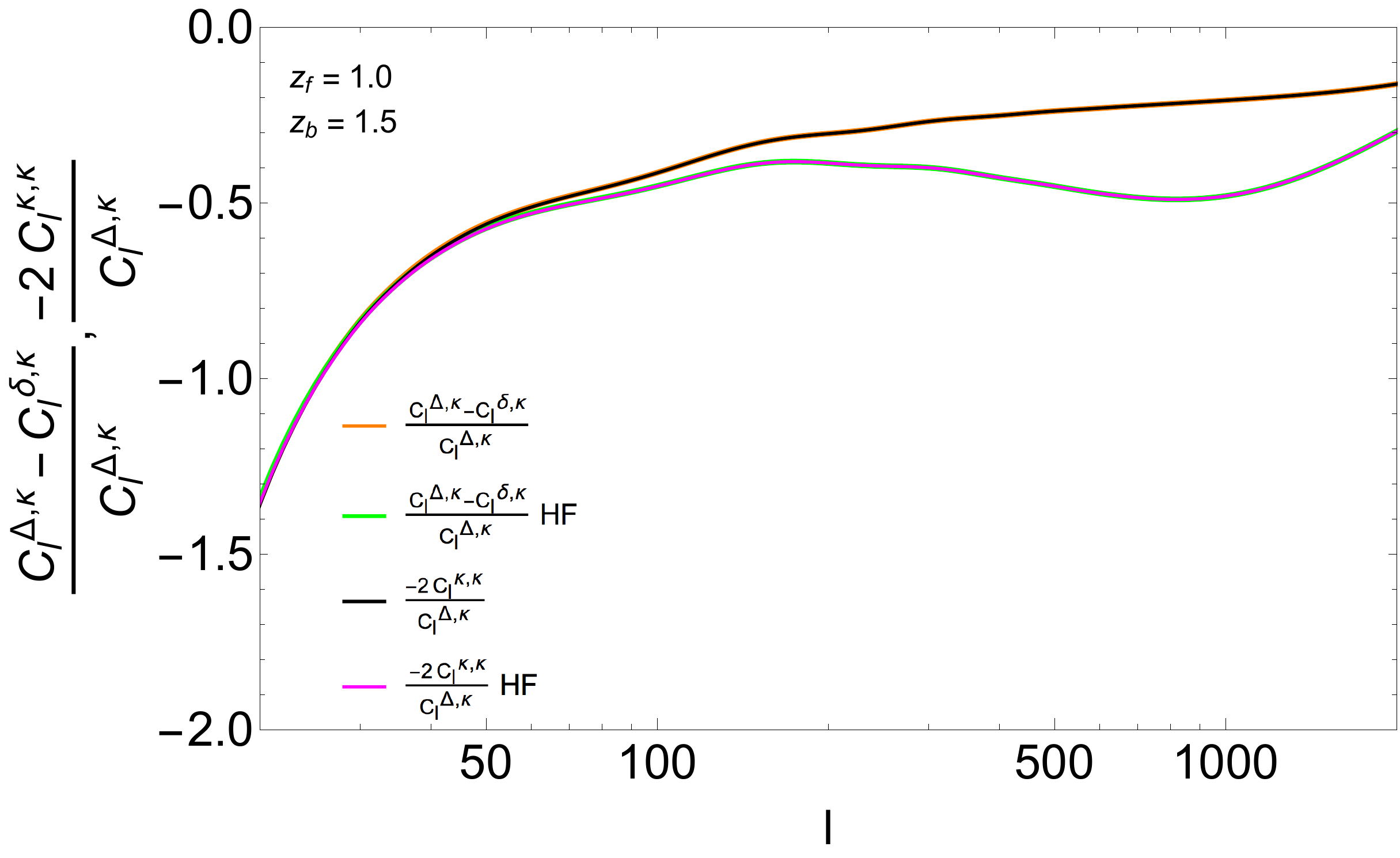}}\\
 {\includegraphics[width=0.8\textwidth]{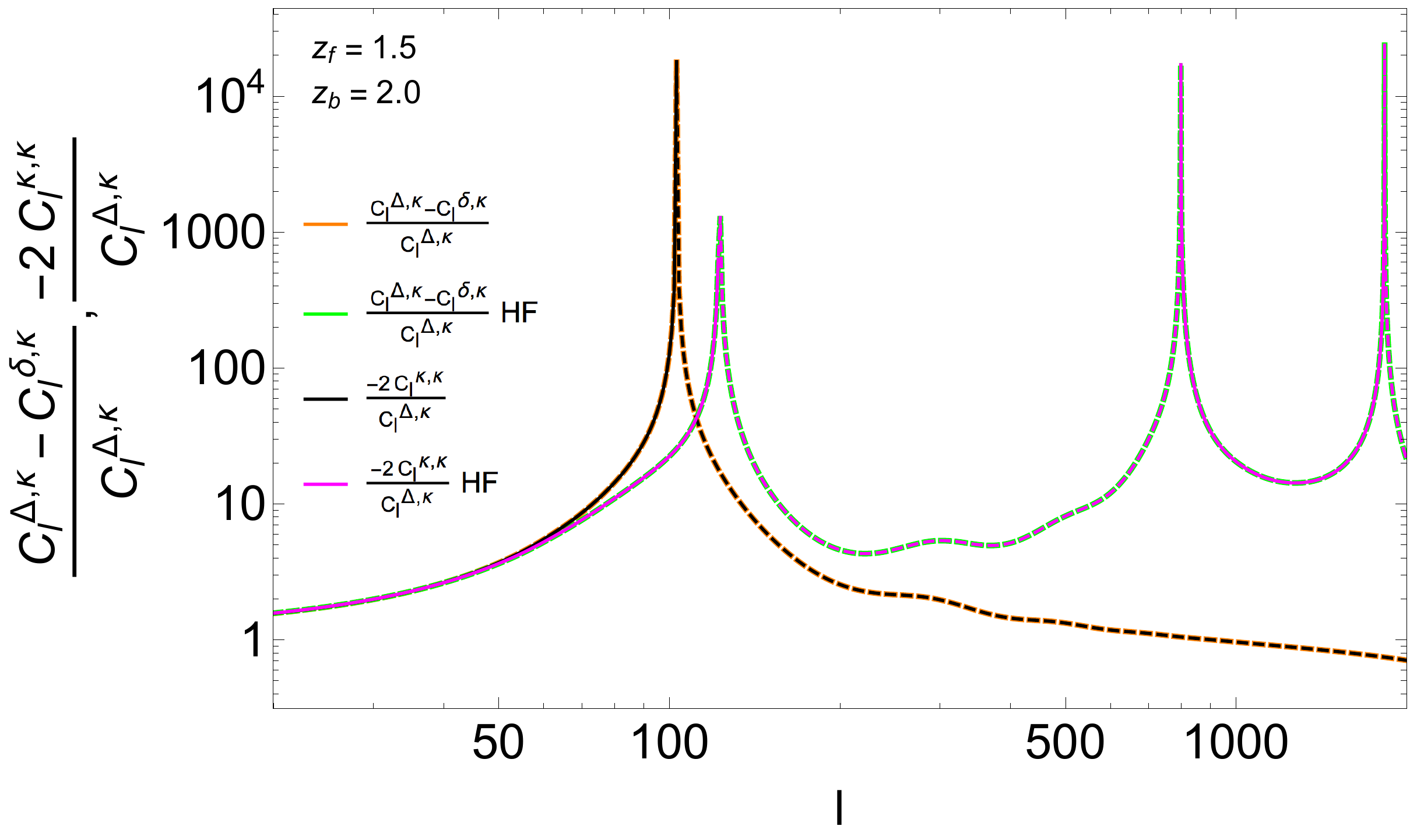}}
\caption{The top and bottom panels show the relative difference of the full number counts which include relativistic corrections, and the contribution from only $\delta$ to the number count-tangential shear correlation (orange for linear and green for Halofit). The contribution of lensing $\kappa$ alone is also shown (black for linear and magenta for Halofit). For large $\ell$ the full difference and the $\ka-\ka$ term are identical while at low $\ell$ there is no difference between the linear perturbation theory result and Halofit. We consider the cases $z_f=1.0$ and $z_b=1.5$, and $z_f=1.5$ and $z_b=2.0$ respectively. The dashed lines in the bottom panel indicate negative signals. The spikes are due to sign changes.}
 \label{f:highz}
\end{figure}

As our numerical examples illustrate, the impact of especially lensing on the number counts increases with redshift.
Future surveys like Euclid~\cite{Amendola:2016saw,EuclidRB,EuclidWeb},  LSST (Large Synoptic Survey Telescope)~\cite{Abell:2009aa,LSSTWeb} or SKA (Square Kilometer Array)~\cite{Abdalla:2015zra,Maartens:2015mra} will go to higher redshifts than the here modeled DES-like redshift binning, and it will be even more important to take lensing effects on the number counts into consideration. In the case of SKA, the lensing term is of course absent for intensity maps~\cite{Hall:2012wd} but present for number counts. As an example, in Fig.~\ref{f:highz}, we show the situation for two cases: $z_f=1.0$, $z_b=1.5$ and $z_f=1.5$, $z_b=2.0$. In both cases, the $\kappa-\kappa$ contribution is seen to be identical to the full difference $(C_\ell^{\De,\ka}-C_\ell^{\de,\ka})$, and is therefore evidently the most significant effect.

\section{Conclusion}\label{s:con}
Galaxy surveys are these days analyzed via mainly two approaches, one of which is to investigate galaxy clustering and one is to detect gravitational shear, also known as gravitational lensing. It has recently become popular to cross-correlate these two observations, such that the correlations between the number density in a foreground bin, and the shear in a background bin are measured. In a relativistic framework, galaxy catalogues do however not measure purely the density $\de$ but the combination $\De$ of density, redshift   space distortions, lensing and large scale relativistic terms as given in Eq.~(\ref{e:DezNF}). We have investigated this issue in this paper.

The number density of sources in a foreground bin is, in a relativistic setting, not a direct observable. Most prominently, the number density is affected by redshift space distortions and by lensing itself, but also by the large scale relativistic effects and the Doppler term. These introduce additional correlations with a lensing bin in the background. Owing to these extra-correlations, it is not ideal to simply `combine probes' as is frequently done, and to compute a joint covariance matrix for lensing, galaxy counts and galaxy-galaxy lensing. Rather, the relativistic corrections in the number counts should be directly accounted for as signal. In this paper we have shown that for a DES-like redshift binning, this can lead to a 50\% correction of the signal in the  density-tangential shear correlation function of the highest redshift bin. For the lowest redshift bin, the correction due to relativistic effects is 1.5\%. This contribution systematically enhances the correlation $\De$-$\ka$. In our treatment we have set $s=0$ which is probably not a very good approximation especially for the highest redshift bins. Including the correct value for $s(z)$ (which we do not know) will reduce the correction somewhat. 

Not including the relativistic effects in the signal, but computing a joint covariance matrix and marginalizing over galaxy bias parameters, will hide the relativistic corrections in the error bars and in the marginalized biases. This is a sub-optimal procedure from a theoretical perspective, as galaxy clustering already occurs in Newtonian gravity, but the here discussed relativistic effects are a signal of General Relativity, and hence contribute to our physical understanding of the Universe. For future surveys with decreasing errors and higher redshifts, not hiding these effects in the error bars will also be important in order to reach the targeted percent accuracy on cosmological parameters. With increasing sky coverage and redshift, the relativistic effects can even dominate the cross-correlation on large angular scales.

Note also, that the claim that lensing be relevant only on very small scales is simply not correct. Its relative contribution to the the total number-count-tangential shear power spectrum is nearly constant from $\ell\sim 50$ to $\ell=2000$ with a wide hump around $\ell\sim 700$. The increase of lensing at smaller scales is therefore similar to the one of density fluctuations.

The goal of this brief and simple study is not a detailed signal to noise analysis of the effect in the DES data. It is possible that in the highest redshift bin the density-tangential shear cross correlation in the first-year analysis of the DES data is systematically biased by a factor of more than 50\% so that our correction would be smaller than the error. We also have not analysed how neglecting lensing in these cross-correlations propagates into the parameter estimation from DES. This would require a more detailed study taking into account also the number count spectra where lensing also is not considered in the present DES data analysis. Nevertheless, we think such a large effect has to be at least discussed and, as we show here, it is relatively easy to include it. A more detailed signal to noise analysis of the effect in the DES data is left for a future work, maybe in collaboration with DES. 

Finally, we want to stress that 
including the $\ka-\ka$ term in the analysis is not only necessary but also very fruitful. This term is sensitive to the lensing potential and it contains additional information which we can use, e.g., to test modified gravity models, see~\cite{Montanari:2011nz}.

\section*{Acknowledgement}
 It is a pleasure to thank Jonathan Blazek, Daniel Gruen, Stefan Hilbert, Elisabeth Krause and Bastian Muellerthann for scientific discussions and background information. We are grateful to Anthony Challinor for pointing out an error in the first draft.
This work is supported by the Swiss National Science Foundation.\vspace{2cm}

\appendix

\section{Deriving the correlation function for number counts and tangential shear}
\label{a:der}
In this appendix we derive the correlation between density number counts and the tangential shear and we make contact with the formulas usually found in the literature.

\subsection{Full Sky}
\begin{figure}[h!]
\begin{center}
\includegraphics[width=\textwidth]{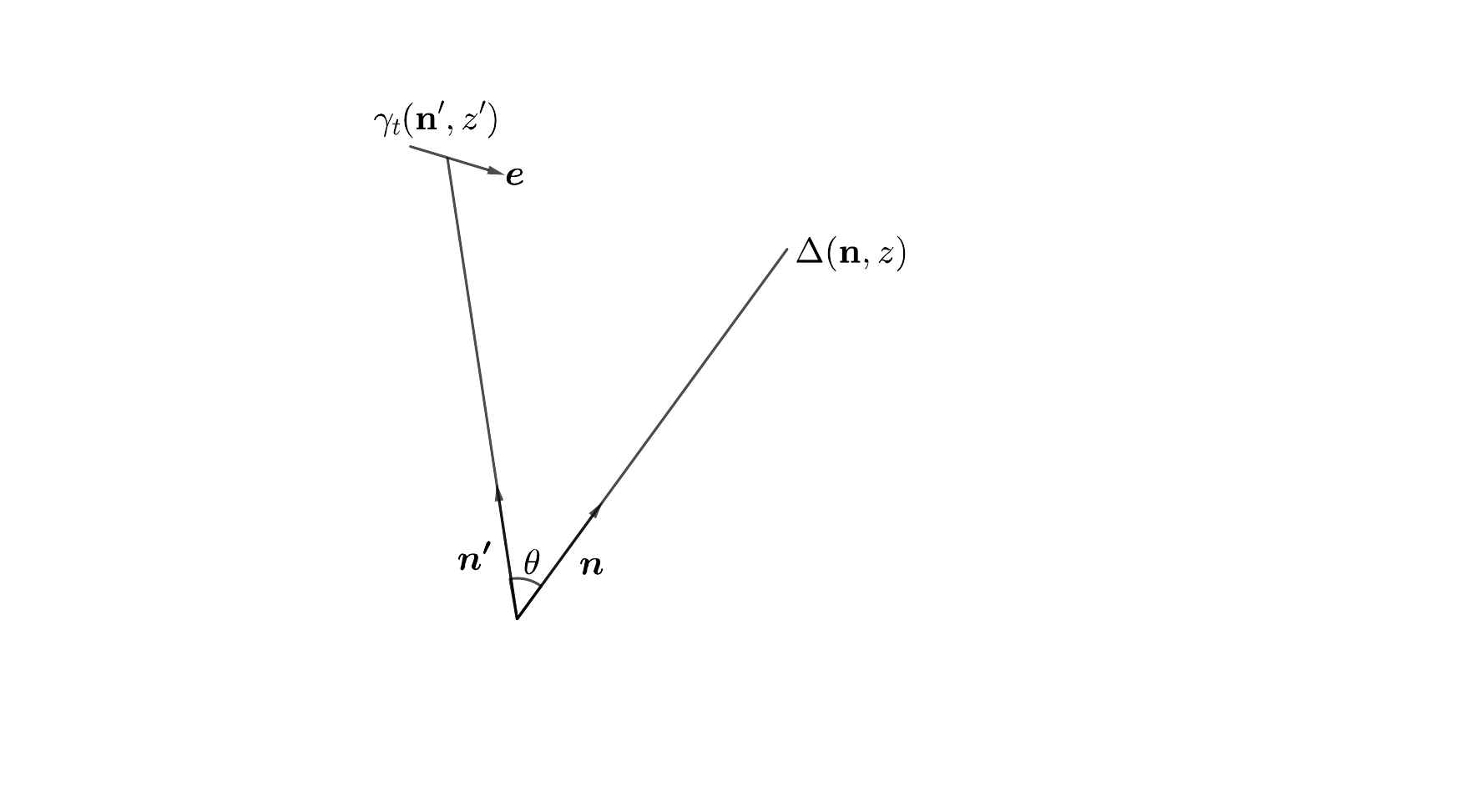}
\end{center}
\caption{\label{f:geo}Notation for the correlation between the galaxy number density fluctuation $\De(\bn,z)$ and the tangential shear $\ga_t(\bn',z')$ in direction $\bfe$.}
\end{figure}

We consider the correlation between the galaxy number density in direction $\bn$ at redshift $z$, $\De(\bn,z)$ and the tangential shear at $(\bn',z')$ perpendicular to $\bn'$ in direction $\bfe$ which points from $\bn'$ towards $\bn$, see Fig.~\ref{f:geo}. 
The $2\times 2$ shear tensor $\gamma_{ab}$ is given in terms of the lensing potential $\phi$ by
\bea
\ga_{a b}(\bn',z') &=&- (\nabla_a\nabla_b-\frac{1}{2}\de_{ab}\De_\Om)\phi  \\
\ga_t(\bn',z') &=&  \ga_{a b}(\bn',z') e^ae^b \,,
\eea
where $\delta_{ab}$ is the $2\times 2$ identity matrix, $\nabla_a$ is the covariant derivative on the sphere and $\bfe$ is the tangent vector on the sphere pointing from $\bn'$ to $\bn$. 
We expand the fields $\De$ and $\phi$ in spherical harmonics,
\bea
\phi(\bn',z') &=& \sum_{\ell,m}a^\phi_{\ell m}(z')Y_{\ell m}(\bn')\\
\De(\bn,z) &=& \sum_{\ell,m}a^\De_{\ell m}(z)Y_{\ell m}(\bn) \,.
\eea
Without loss of generality we may choose $\bn=\bfe_z$ in $z-$direction and $\bfe=-\bfe_\vth$ in $-\vth$-direction. Using that $Y_{\ell m}(\bfe_z)=\de_{m0}\sqrt{\frac{2\ell+1}{4\pi}}$ we then obtain
\bea
\langle\De(\bn,z)\ga_t(\bn',z')\rangle &=& -\hspace{-3.5mm}\sum_{\ell,m,\ell',m'}\langle a^{\De\, *}_{\ell m}(z)a^\phi_{\ell' m'}(z')\rangle Y^*_{\ell m}(\bfe_z)\left(\nabla_\vth\nabla_\vth-\frac{1}{2}\De_{\Om}\right) Y_{\ell' m'}(\bn')   \nonumber\\
 &=& -\sum_\ell C_\ell^{\De,\phi}(z,z')\sqrt{\frac{2\ell+1}{4\pi}}\left(\nabla_\vth\nabla_\vth-\frac{1}{2}\De_{\Om}\right) Y_{\ell 0}(\bn') \,.  \label{ea:cor}
\eea
Here we assumed that the fluctuations $\De$ and $\phi$ are statistically isotropic such that
\be\label{ea:5}
\langle a^{\De\, *}_{\ell m}(z)a^\phi_{\ell' m'}(z')\rangle =  C_\ell^{\De,\phi}(z,z')\de_{\ell,\ell'}\de_{m,m'}  \,.
\ee
Next we use the spin raising and spin lowering operators (see~\cite{Durrer:2008aa}, Appendix~4 for details) to write the covariant
derivative
$\nabla_\vartheta = \frac{1}{2}(\spart + \spart^*)$ and $\De_\Om =\frac{1}{2}(\spart\spart^* + \spart^*\spart)$ so that
$$
\nabla_\vth\nabla_\vth-\frac{1}{2}\De_{\Om} =\frac{1}{4}\left(\spart^2+\spart^{*\,2}\right)
$$
Now in terms of $\mu=\cos\vth'$ for a function (spin $s=0$) which does not depend on $\varphi$ we find (again, details are found in ~\cite{Durrer:2008aa}, Appendix~4)
\bean
\spart^2f(\mu) &=& \spart^{*\,2}f(\mu) ~=~ (1-\mu^2)f''(\mu)  \,.
\eean
Inserting this in $Y_{\ell 0}(\bn')=\sqrt{\frac{2\ell+1}{4\pi}}P_\ell(\mu)$ we obtain
\bean\frac{1}{4}\left(\spart^2+\spart^{*\,2}\right)Y_{\ell 0}(\bn') &=& \frac{1}{2}\sqrt{\frac{2\ell+1}{4\pi}}(1-\mu^2)P''_\ell(\mu) \\
&=& \frac{1}{2}\sqrt{\frac{2\ell+1}{4\pi}}P_{\ell\,2}(\mu)\,.
\eean
Here $P_{\ell\,2}$ is the associated Legendre function of order $2$ (see~\cite{Abramowitz:1972} and \cite{Durrer:2008aa}, Appendix~4 for details).
Inserting this in \eqref{ea:5} we obtain
\bea
\langle\De(\bn,z)\ga_t(\bn',z')\rangle &=&\frac{-1}{8\pi}\sum_\ell C_\ell^{\De,\phi}(z,z')(2\ell+1)P_{\ell\,2}(\cos\vth') \\
&=&\frac{-1}{4\pi}\sum_\ell C_\ell^{\De,\ka}(z,z')\frac{2\ell+1}{\ell(\ell+1)}P_{\ell\,2}(\bn\cd\bn') \,, 
\label{ea:corl2}
\eea
where we have used that $a^\ka_{\ell m}=(\ell+1)\ell a^\phi_{\ell m}/2$.
This result implies that the corresponding correlation spectrum is
\be
C_\ell^{\De,\ga_t}(z,z')=-C_\ell^{\De,\ka}(z,z')  \,.
\ee
The angular dependence via $P_{\ell\,2}(\bn\cd\bn')$, is a consequence of the fact that we are correlating the 2-tensor $\ga_{ab}$ with a scalar quantity, hence the corresponding correlation function $\langle\De(\bn,z)\ga_t(\bn',z')\rangle$ transforms like a tensor under rotations around $\bn'$.
 
This remains true for the correlation of an arbitrary scalar quantity $A$ with the tangential shear which is therefore given by \eqref{ea:corl2} replacing $C_\ell^{\De,\ka}(z,z')$ by $C_\ell^{A,\ka}(z,z')$.  Note also that the normalisation scales with $\ell$  as expected since
 $$
 \int_{-1}^1P_{\ell\,s}(\mu)P_{\ell'\,s}(\mu)d\mu = \frac{2}{2\ell+1}\frac{(\ell+s)!}{(\ell-s)!}\de_{\ell,\ell'}\,. 
 $$
 
\subsection{Flat sky}

Usually the above equation is derived somewhat differently in the flat sky approximation. This is largely sufficient if one considers relatively small sky patches as e.g.~the DES year-1 data with their $\approx$1300 square degrees. In flat sky $\bel$ is a 2d vector, the Fourier transform variable of the sky position $\bx$ which is dimensionless. The spherical harmonics are then replaced by $Y_{\ell m}\ra \frac{1}{2\pi}\exp(i\bel\cdot\bx)$, $\spart = -\dd_1 +i\dd_2$ and $\spart^* = -\dd_1 -i\dd_2$.
One can obtain the flat sky result of the above equation directly by using the flat sky versions of $\spart$ and $\spart^*$. This yields 
\be\label{ea:ddflat}
\frac{1}{2}(\spart^2 + \spart^{*\,2})Y_{\ell m} ~\ra~ -\frac{\ell^2}{4\pi}\left(\cos^2\varphi-\sin^2\varphi\right)e^{i\bel\cdot\bx} =\frac{\ell^2}{2\pi}\left(\frac{1}{2}-\cos^2\varphi\right)e^{i\bel\cdot\bx}
\ee
Statistical isotropy in the flat sky yields
$$\langle\De^*(\bel,z)\phi(\bel',z')\rangle=\de^2(\bel-\bel')C^{\De,\phi}_\ell(z,z'),$$ 
and the convergence is given by
$\ka(\bel) = \frac{\ell^2}{2}\phi(\bel) $. 
Note that correctly speaking we consider two flat skies, one at redshift $z$ where the foreground galaxies lie and one at redshift $z'$ where we measure the shear of the background galaxies.

Using $\cos\varphi = \bel'\cdot\bfe/\ell'$, \eqref{ea:ddflat} leads to
\be
\ga_t(\bx,z') =\frac{1}{2\pi}\int d^2\bel' \left((\bel'\cd\bfe)^2-\frac{1}{2}\ell'^2\right)e^{-i\bel'\cdot\bx}\phi(\bel',z'), 
\ee
so that we  obtain for the correlation function
\bea
\langle\De(\by,z)\ga_t(\bx,z')\rangle &=&
\frac{2}{(2\pi)^2}\int d^2\ell \left((\hat\bel\cd\bfe)^2-\frac{1}{2}\right)e^{ir\bel\cd\bfe} C^{\De,\ka}_\ell(z,z') \,.
\eea
In the last line we set $\hat\bel= \bel/\ell$ and we have used statistical isotropy and $\ka(\bel)=\ell^2\phi(\bel)/2$.
The angular integration gives
\be
\int_0^{2\pi}\left(\cos^2\varphi-\frac{1}{2}\right)e^{ir\ell\cos(\varphi)}d\varphi = -\pi J_2(r\ell) \,,
\ee
where $J_2$ is the Bessel function~\cite{Abramowitz:1972} of order 2. With this we find
\be
\langle\De(\bx,z)\ga_t(\bx+\br,z')\rangle =-\frac{1}{2\pi}\int_0^\infty \ell d\ell J_2(\ell r)C_\ell^{\De,\ka}(z,z') \,.
\ee
To make contact with the formulas found in the literature, e.g. in~\cite{Abbott:2017wau}, 
we consider a distribution $n_f(z)$ of foreground galaxies and a distribution $n_b(z')$ of background shear measurements. Integrating over these distributions and using that a foreground galaxy at $z$ only shears a background one at $z'$ for $z<z'$, we obtain
\be
\langle\De^{(f)}(\bx)\ga^{(b)}_t(\bx+\br)\rangle =\frac{-1}{2\pi}\int_0^\infty dz n_f(z)\int _z^\infty dz' n_b(z')\int_0^\infty \ell d\ell J_2(\ell r)C_\ell^{\De,\ka}(z,z') \,.
\ee
The difference of this expression from the one usually used in the literature is that there the approximation $C_\ell^{\De,\ka}(z,z') \sim C_\ell^{\de,\ka}(z,z')$ is used, i.e.~it is assumed the number density of foreground objects is (modulo a bias factor) given by the underlying density $\delta$. One can then express the shear power spectrum as an integral over the gravitational potential which is related to the density by the Poisson equation. Using the Limber approximation one can write the result as an integral over the dimensionless  power spectrum  in $k$-space. This leads after some standard manipulations to
\be
\langle\de\ga_t\rangle(\theta) =\frac{3\Om_mH_0^2}{2}\int_0^\infty \frac{\ell d\ell}{2\pi} J_2(\ell \theta)\int_0^\infty \!\!dz n_f(z)\int _z^\infty \!\!dz' n_b(z')\frac{r(z)(r(z')-r(z))}{r(z')H(z)}P\left(\frac{\ell+1/2}{r(z)},z\right) \,.
\ee
This is the expression found, e.g.,  in~~\cite{Abbott:2017wau} 
where for $n_f(z)$ and $n_b(z')$ we have to consider the distribution of foreground respectively background galaxies in the different redshift bins. Here, $P$ is the (dimensionless) Fourier space density fluctuations spectrum, and in our numerical applications we have used Halofit~\cite{Takahashi:2012em} to model its non-linearities.
\vspace*{2cm}

\bibliography{refs}
\bibliographystyle{JHEP}
\end{document}